\documentclass[a4paper, aps, pra,twocolumn]{revtex4-1}

\usepackage[english]{babel}
\usepackage{amsmath}
\usepackage{mathtools}
\usepackage{graphicx}
\usepackage{hyperref}
\usepackage{braket}
\usepackage[colorinlistoftodos]{todonotes}

\newcommand{\dket}[1]{\left.\Ket{#1}\right\rangle}
\newcommand{\dbra}[1]{\left\langle\Bra{#1}\right.}
\newcommand{\dbraket}[1]{\left\langle\Braket{#1}\right\rangle}
\newcommand{\rSA}{\rho_R^A}

\let\Re\relax \DeclareMathOperator\Re{\mathrm{Re}}%
\let\Im\relax \DeclareMathOperator\Im{\mathrm{Im}}%
\DeclareMathOperator\Tr{\mathrm{Tr}}%
\DeclareMathOperator\sinc{\mathrm{sinc}}%

\begin{document}
\title{Efficient Real-Time Path Integrals for Non-Markovian Spin-Boson Models}

\author{A.~Strathearn}
\affiliation{SUPA, School of Physics and Astronomy, University of St Andrews, St Andrews, KY16 9SS, United Kingdom}

\author{ B.~W.~Lovett}
\email{bwl4@st-andrews.ac.uk}
\affiliation{SUPA, School of Physics and Astronomy, University of St Andrews, St Andrews, KY16 9SS, United Kingdom}

\author{ P.~Kirton}
\affiliation{SUPA, School of Physics and Astronomy, University of St Andrews, St Andrews, KY16 9SS, United Kingdom}

\date{\today}

\begin{abstract}

Strong coupling between a system and its environment leads to the emergence of non-Markovian dynamics, which cannot be described by a time-local master equation. One way to capture such dynamics is to use numerical real-time path integrals, where assuming a finite bath memory time enables manageable simulation scaling. However, by comparing to the exactly soluble independent boson model, we show that the presence of transient negative decay rates in the exact dynamics can result in simulations with unphysical exponential growth of density matrix elements when the finite memory approximation is used.
We therefore reformulate this approximation in such a way that the exact dynamics are reproduced identically and then apply our new method to the spin-boson model with superohmic environmental coupling, commonly used to model phonon environments, but which cannot be solved exactly.
Our new method allows us to easily access parameter regimes where we find revivals in population dynamics which are due to non-Markovian backflow of information from the bath to the system.

\end{abstract}

\maketitle

\section{Introduction}

Finding accurate descriptions of open quantum systems strongly coupled to external environments  is essential for understanding how quantum systems lose their coherence.~\cite{decoherencereview, breuer_petruccione_2002}. For example, the behaviour of quantum dots in semiconductors~\cite{QDstrongcavity,QDdamping, QDlambshift, QDstrongexphon, daranonmarkov} and NV centres in diamonds~\cite{NVspincoherence,NVdecohspinbath} can both require a strong coupling description. Strong coupling inevitably leads to the emergence of non-Markovian phenomena in such systems, and this has been experimentally demonstrated~\cite{liu_nonmarkexp_2011}, opening up the potential for exploiting non-Markovianity as a resource in developing quantum technologies~\cite{nonmark_quantchan, huelga_rivas_plenio_2012}.

In this paper we will introduce a new way of modelling a strongly coupled environment, developing previous approaches based on Feynman's path integral formulation of open quantum systems~\cite{feynman_vernon_1963, feynman_hibbs_1965, caldeira_leggett_1983, weiss_2012}. In particular, we will show how a technique based on discretisation of the Feynman influence functional -- the so-called Augmented Density Tensor (ADT) can be modified to significantly improve the convergence of simulations in its numerical implementation. This allows us to study 
the spin-boson model in a very strong coupling regime that shows clear non-Markovian behaviour that we quantify with the widely-used trace distance measure~\cite{blpmeasure}.

When the system-environment coupling is weak, Born-Markov master equations~\cite{lindblad_1976, breuer_petruccione_2002} provide a perturbative approach in which the environment is assumed to be Markovian i.e.\ memoryless.
While this approach is very successful within its range of applicability~\cite{breuer_petruccione_2002, nazir_mccutcheon_2016}, for many physical applications the (rather severe) approximations made when deriving these types of master equation are not justified.
A common approach to go beyond the weak coupling regime is to use a polaron-transformed master equation~\cite{mccutcheon11, royhughestriplet, pollock_polaron_2013, nazir_mccutcheon_2016, jang2008theory, jang2009theory}.
Such an approach has been used to understand features which could not be explained in weak coupling for semiconducting quantum dots~\cite{mccutcheon11}, circuit QED~\cite{marthaler2011}, energy transfer in biological systems~\cite{kolli12, zimanyi2012, pollock_polaron_2013} and Bose-Einstein condensation of photons in optical microcavities~\cite{kirton2013, kirton2015}. However, a polaron master equation comes at the expense of introducing a restriction on the renormalized system Hamiltonian terms. Other master equation based techniques such as Nakajima-Zwanzig projection operator equations~\cite{breuer_petruccione_2002}, time-convolutionless master equations~\cite{breuer_petruccione_2002}, reaction co-ordinate methods~\cite{IlesSmith2014} and Keldysh-Lindbald equations~\cite{Muller2017} are able to pertubatively go beyond both the Born and Markov approximations but are still, in practice, limited to restricted parameter regimes.

One can overcome this restriction by using non-perturbative approaches. One class of such methods uses the Feynman path integral representation and formally integrates out the environmental degrees of freedom. Then all effects of the system-environment coupling are described by an influence functional~\cite{feynman_vernon_1963} that acts only on the reduced system trajectories.
This representation of an open quantum system has proved useful in developing both analytical and numerical methods.
As an example, the non-interacting blip approximation~\cite{caldeira_leggett_1983} is a successful non-perturbative analytical method for understanding the spin-boson model, derived from the influence functional. However, this is only applicable for a relatively small range of parameters.

A versatile numerical method for finding the dynamics of these kinds of systems is the ADT scheme, first introduced as the Quasi-Adiabatic Propagator Path Integral (QUAPI) by~\citet{makri_makarov_1995_i,makri_makarov_1995_ii}. This is, in principle, applicable to environments with arbitrary spectral density.
The main drawback to this technique is the exponential scaling of storage requirements with system size, though in some cases ways of reducing these requirements are known~\cite{sim_2001,makri_2017}.
Hence, the primary use of this method is to calculate the of dynamics of few-level systems in contact with a bosonic reservoir. 
The ADT has been used to calculate equilibrium correlation functions \cite{shao_makri_2002} and system steady states~\cite{makri_makarov_1995_i}. In addition, multiple spatially separated sites coupled to the same bath can be accounted for~\cite{nalbach_eckel_thorwart_2010}, and additional Markovian dynamics in the reduced system can be incorporated at no additional computational cost~\cite{axt_2016}.
Practically, the ADT method has been put to use effectively in, for example, benchmarking master equation methods~\cite{chang_zhang_cheng_2013,variational_quapi}, simulating systems in the difficult-to-access regime of subohmic system-bath coupling~\cite{nalbach_thorwart_2010} and investigating dissipative dynamics of charge qubits in realistic environments~\cite{thorwart_2005,eckel_weiss_thorwart_2006}. Very recently, it was shown that, as long as the operator which couples to the environment only acts in a small section of Hilbert space much larger systems can be treated at only small numerical cost~\cite{cygorek2017}.

Other numerical approaches are able to accurately describe non-perturbative dynamics in spin-boson systems. The hierarchical equations of motion method~\cite{tanimura_kubo_1989} is regularly used to benchmark master equation approaches~\cite{IlesSmith2014}. Techniques based on the numerical renormalisation group have been successfully applied to quantum impurity dynamics by mapping baths characterized by particular spectral density forms into semi-infinite one-dimensional chains~\cite{anders05,Bulla2005, anders07,bulla08}. Ansatz wavefunctions such as matrix product states~\cite{yaoyao13} can also be used following a similar mapping~\cite{chin10}, while approaches based on a Monte Carlo sampling of the path integral can be used to look at dynamics~\cite{Egger1994, Kast2013}.  For a recent review of many of the techniques mentioned above see \citet{deVega2017}. 

The key approximation made in the ADT scheme is that the bath is assumed to have a sharply defined finite memory time,
and so non-Markovian effects are given finite range. This is intuitively justified since, for infinite bosonic baths, all correlations decay to negligible size in finite time.
The performance of this method has been successfully tested against other numerical methods~\cite{nalbach_ishizaki_fleming_thorwart_2011} and exact analytics~\cite{thorwart_reimann_hanggi_2000}. However, the full consequences of the sharp memory time cutoff and how this affects convergence have not been addressed. Moreover, it has been recognized that for superohmic environments using this approximation results in both quantitatively and qualitatively incorrect asymptotic long time behaviour~\cite{quapi_so1,quapi_so2}. 
In this paper we will show that, in certain cases, throwing away small long time correlations beyond the cutoff can have dramatic effects, including unbounded growth of density matrix elements. We then propose a less severe way to make this approximation which does not suffer from the same problems. 

The structure of this paper is as follows: In Sec.~\ref{sec:scheme} we introduce the ADT scheme in detail. Following this, in Sec.~\ref{sec:indbos}, we investigate the results of enforcing the finite memory approximation upon the exact solution of a pure dephasing model and identify the class of bath spectral densities for which the finite memory approximation has most drastic effects. We then propose an alternative way of taking the finite memory approximation such that the exact solution of the dephasing model is reproduced at all times for arbitrary spectral densities. In Sec.~\ref{sec:spinbos} we go on to apply our method to the spin-boson model with a highly non-Markovian superohmic bath. We find that our modified memory cutoff significantly improves convergence in this model and allows us to observe revivals in the population dynamics due to excitation exchange with the environment. We also calculate the trace distance measure of non-Markovianity in this model, finding regimes of non-Markovian behaviour that we compare to the population revivals. Finally, in Sec.~\ref{sec:summary} we summarise our results.

\section{ADT Scheme}
\label{sec:scheme}

Before introducing our improved memory cutoff, we first review the ADT scheme and the approximations required. This will also aid us in introducing our notation.
The class of models to which the scheme is applicable are those consisting of a small system of interest linearly coupled to a macroscopic bath of bosonic modes. The generic Hamiltonian of such models is
\begin{align}
\label{eq:hamil}
 H=&H_0+ \hat{s}\sum_\alpha \hat{B}_\alpha + \sum_\alpha \omega_\alpha a^\dagger_\alpha a_\alpha \\
=&H_0+H_B,
 \end{align}
where $H_0$ is the free system Hamiltonian and $H_B$ contains both the free bath Hamiltonian and the system-bath interaction. Here,  $a_\alpha^\dagger$ ($a_\alpha$) and $\omega_\alpha$ are the creation (annihilation) operators and frequencies of the $\alpha$th oscillator. The system operator $\hat{s}$ couples to the bath operators $\hat{B}_\alpha=g_\alpha a_\alpha+g_\alpha^* a_\alpha^\dagger$ with coupling constants $g_\alpha$.

To simplify our notation we work in the Liouvillian representation such that operators in Hilbert space are represented by vectors in Liouville space. To parameterise the $D$ dimensional Hilbert space of the system we use the $D$ eigenstates of $\hat{s}$. Operators in this space are vectorized in the following way
\begin{align}
\hat{\rho}_R&=\sum_{s^+, s^-}\rho_{s^+ s^-}\ket{s^+}\bra{s^-} \nonumber\\
&\equiv\sum_{S} \rho_S\dket{S}\equiv\dket{\rho_R},
\end{align}
where the sum over $S$ runs over the $D^2$ pairs of $\{s^+,s^-\}$ and $\hat{s}\ket{s^+}=s^+\ket{s^+}$ and likewise for $s^-$. We use the notation $\dket{x}$ to mean a vector in Liouville space.
The bath Hilbert space can also be represented in a similar way, though we do not need to define its basis explicitly in what follows.
The evolution of the reduced system, assuming factorizing initial conditions is now represented as:
\begin{align}
\label{redrho}
\dket{\rho_R(t)}=\Tr_B\left[\text{e}^{\mathcal{L} t}\dket{\rho_R(0)}\dket{\rho_B}\right],
\end{align}
with the Liouvillian $\mathcal{L}=\mathcal{L}_0+\mathcal{L}_B$, where $\mathcal{L}_0$ and $\mathcal{L}_B$ generate coherent evolution caused by $H_0$ and $H_B$ respectively. Recently it has been shown~\cite{axt_2016} that additional Markovian non-unitary dynamics of the reduced system can be incorporated by adding Lindblad-type dissipators into the free system Louivillian $\mathcal{L}_0$, making it straightforward to account for coupling to other baths for which the Born-Markov approximations are well justified.
In addition to factorising initial conditions we also assume the initial state of the bath is that of thermal equilibrium when no system is present $\rho_B=\exp(-\sum_\alpha \omega_\alpha a^\dagger_\alpha a_\alpha/T)/\mathcal{Z}$, with  temperature $T$ and partition function $\mathcal{Z}$.

The first approximation made to make  Eq.~\eqref{redrho} computable is to factorize the long time propagator into $N$ short time propagators $\text{e}^{\mathcal{L} t}=(\text{e}^{\mathcal{L} \Delta t})^N$ and then to employ a Trotter splitting between the system and bath parts~\cite{trotter1959}
\begin{align}
\label{trot}
\text{e}^{\mathcal{L} \Delta t}\approx \text{e}^{\mathcal{L}_B \Delta t}\text{e}^{\mathcal{L}_0 \Delta t},
\end{align}
on each of these. The error introduced in this process is $\mathcal{O}(\Delta t^2)$.
We note that the argument that now follows can be easily adapted to use a symmetrized Trotter splitting~\cite{suzuki1976,makri_makarov_1995_i,makri_makarov_1995_ii} that improves the error to $\mathcal{O}(\Delta t^3)$. All the numerical results we present do include this symmetrized splitting, but for simplicity of notation we will use the definition in Eq.~\ref{trot} here.

Tracing out the bath degrees of freedom then results in a reduced density matrix at time $t_N=N\Delta t$, whose elements are 
\begin{multline}
\label{discreteevo}
\dbraket{S_N|\rho_R(t_N)}=\sum_{S_0 \ldots S_{N-1}} F\left(\{S_k\}\right) I\left(\{S_k\}\right)\\
\times \dbraket{S_0 |\rho_R(0)} .
\end{multline} 
The functions $F\left(\{S_k\}\right)$ and $I\left(\{S_k\}\right)$ constitute the free part of the evolution and the discretized Feynman influence functional respectively, and are given by
\begin{align}
F\left(\{S_k\}\right)=& \prod_{j=1}^N\dbra{S_j}\text{e}^{\mathcal{L}_0 \Delta t}\dket{S_{j-1}}\\
= & \prod_{j=1}^N\bra{s^+_j}\text{e}^{-iH_0 \Delta t}\ket{s^+_{j-1}}\bra{s^-_{j-1}}\text{e}^{iH_0 \Delta t}\ket{s^-_{j}},\\
\label{discreteif}
I\left(\{S_k\}\right)=&\exp\left(-\sum_{k=1}^N \sum_{k'=1}^k(s^+_k-s^-_k)(\eta_{k-k'}s^+_{k'}-\eta^*_{k-k'}s^-_{k'})\right).
\end{align}
The coefficients $\eta_{k-k'}$ quantify the non-Markovian `interaction' between the reduced system at different times $t_k$ and $t_{k'}$ and are defined as
\begin{equation}
\eta_{k-k'} = \begin{dcases*}
\int_{t_{k-1}}^{t_k}\int_{t_{k'-1}}^{t_{k'}}C(t'-t'')dt''dt' &\text{ $k\ne k'$}\\ 
\int_{t_{k-1}}^{t_k}\int_{t_{k-1}}^{t'}C(t'-t'')dt''dt' &\text{ $k=k'$}
\end{dcases*},
\end{equation}
in terms of the bath autocorrelation function 
\begin{align}\label{bathcorr}
C(t)&=\sum_\alpha \braket{\hat{B}_\alpha(t+s)\hat{B}_\alpha(s)}\\
&=\int_0^\infty d\omega J(\omega)(\coth( \omega/2 T)\cos(\omega t)-i \sin(\omega t)),
\end{align}
where $J(\omega)=\sum_\alpha |g_\alpha|^2 \delta(\omega_\alpha-\omega)$ is the spectral density of the bath.

\begin{figure}
\includegraphics[scale=0.7]{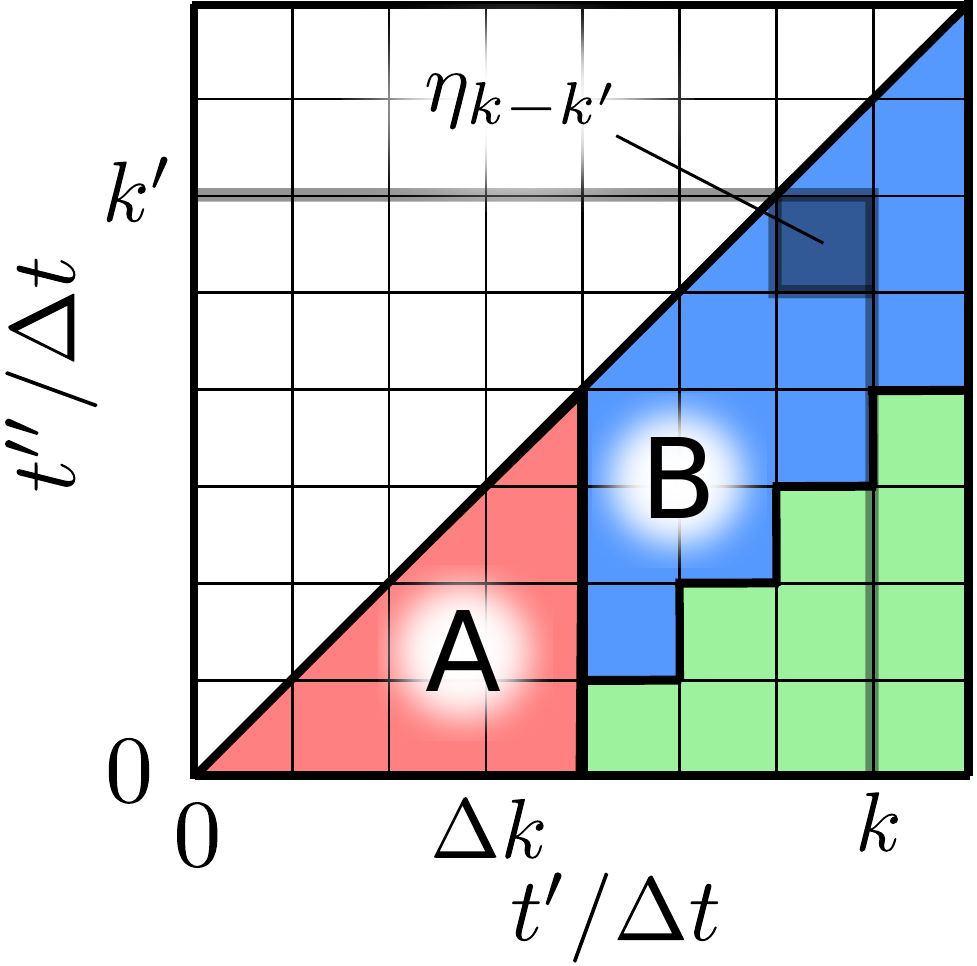}
\caption{A schematic of the region of integration over $C(t)$ appearing in the discrete influence functional, Eq.~\eqref{discreteif}. The exact solution requires all three coloured regions, while making the memory cutoff with $\Delta k=3$ is equivalent to discarding the green region. The $\eta_{k-k'}$ with $k\ne k'$ correspond to the square regions of side $\Delta t$ while for $k=k'$ they are given by the triangular regions along the $t=t'$ line.}
\label{fig:memcut}
\end{figure}

In its current form Eq.~\eqref{discreteevo} is, in principle, numerically computable, though would require the storage of the $D^{2N}$ numbers making up the discretized influence functional.
The exponential dependence of storage requirements on the number of timesteps places strict limits on the length of time for which a given simulation can be propagated.
To circumvent this issue we make use of the \textit{finite memory approximation} originally introduced by~\citet{makri_makarov_1995_i, makri_makarov_1995_ii}.
For a continuum of bath oscillator modes the correlation function Eq.~\eqref{bathcorr} decays at worst algebraically and in finite time becomes effectively zero. This in turn implies that the coefficients $\eta_{k-k'}$ decay to zero as the difference $t_k-t_{k'}$ is increased.
The finite memory approximation involves setting those coefficients for which $k-k'>\Delta k$ such that the time difference $t_k-t_{k'}>\Delta k \Delta t\equiv \tau_c$ identically equal to zero. Here we have introduced $\tau_c$, the bath cutoff time, which must be suitably large to ensure converged results.
In Fig.~\ref{fig:memcut} we visualize this memory cutoff in terms of two dimensional integral regions over $C(t'-t'')$.
For fixed $k$ the coefficients $\eta_{k-k'}$ correspond to columns of a number $\Delta k$ square integral regions extending up to the triangular regions which give $\eta_0$. The finite memory approximation is then made by setting all of the coefficients in the green region to zero. This has the effect that the dynamics we calculate in region A are exact and the approximation only effects what happens for times $t>\tau_c+\Delta t$.

With the finite memory approximation we can now reformulate Eq.~\eqref{discreteevo} in terms of the iterative propagation of an object known as the augmented density tensor (ADT as before).
This propagation is computationally realised most efficiently as contractions between multidimensional tensors but the analogy to standard density matrix propagation is most easily seen by representing it as operator-vector multiplication in an augmented Liouville space:
\begin{align}
\label{eq:quapiprop}
\dket{\rSA(t+\tau_c)} = \text{e}^{\mathcal{L}^A \tau_c} \dket{\rSA(t)},
\end{align}
where $\dket{\rSA}$ is the vectorized ADT and $\mathcal{L}^A$ is the augmented Liouvillian.
The augmented space is constructed by taking the product of $\Delta k$ copies of the original Hilbert space.
The basis of this space at a given time $t_k$ is constructed by taking a product of reduced system basis states at that time and the previous $\Delta k -1$ times $t_{k-1} \ldots t_{k-\Delta k+1}$
\begin{equation}\label{eq:lbasis}
\dket{S^A_{k}} = \dket{S_{k}} \dket{S_{k-1}} \ldots \dket{S_{k-\Delta k+1}},
\end{equation}
thus giving the augmented Hilbert space a dimension of $D^{2 \Delta k}$.
The information stored in the ADT is a set of amplitudes weighting each of the trajectories the reduced system could have taken through its Hilbert space in the previous $\Delta k$ timesteps of the evolution.
The physical reduced density matrix at a given time is then found by summing these amplitudes:
\begin{equation}
\dket{\rho_R(t_k)} = \sum_{S_{k-1}\ldots S_{k-\Delta k+1}} \dbra{S_{k-1}} \ldots \dbraket{S_{k-\Delta k +1}| \rSA (t_k)}.
\end{equation}

Computing Eq.~\eqref{discreteevo} with the memory cutoff requires the storage of the $D^{2 \Delta k}$ elements of the vector representing the augmented state and the $D^{4 \Delta k}$ elements of the propagator matrix.
As mentioned above this is not the optimal representation for actually carrying out the propagation; in fact the alternative tensor representation of the propagator only has $D^{2( \Delta k+1)}$ elements.
In Appendix~\ref{app:prop} we explicitly construct this more efficient tensor propagation and also give the matrix elements of the propagator, $\dbra{S^A_{k}}e^{\mathcal{L}_A \tau_c} \dket{S^A_{k-\Delta k}}$, as well as the components of the initial augmented state vector, $\dbraket{S^A_{\Delta k}|\rSA(\tau_c)}$, required for both tensor and matrix representations of the method.

The scheme then has two parameters which need to be adjusted to ensure convergence of results. 
The memory cutoff time, $\tau_c$, should be made large enough, by increasing $\Delta k$, such that increasing it any further has negligible effects on the final result.
At the same time the error induced by the Trotter splitting must also be eliminated by decreasing the timestep $\Delta t$, which in turn decreases $\tau_c$ for a given $\Delta k$.
Thus, achieving the best results requires both maintaining a small enough timestep $\Delta t$ to eliminate the Trotter error while at the same time keeping it large enough such that the number of timesteps required to capture the correlation time of the bath is kept as small as possible.
Details on the method for finding the optimal value of $\Delta t$ that minimises the overall error can be found in Ref.~\cite{eckel_weiss_thorwart_2006}.

Finally, we note that the ADT has many properties in common with a standard density matrix: it is both Hermitian and has unit trace, from which it follows that a system density matrix calculated from it is also guaranteed to have these properties.
To describe a physical state it is also necessary that the system density matrix is positive. Proving this is, in general, more difficult and it is known that the ADT scheme does not guarantee positivity of the reduced system state~\cite{quapi_so1}.
In the following section we will gain an understanding of how non-positive reduced system states can occur and describe a procedure to help prevent this from happening.

\section{Memory Cutoff in an Exactly Soluble Model}
\label{sec:indbos}

To investigate the role of the finite memory approximation in the occurrence of non-positive reduced system states, here we study the effects of making this approximation on an exactly soluble pure dephasing model where $[H_0,\hat{s}]=0$.
The simplest example of such a model is that of a two level system described by the Hamiltonian $H_0=\epsilon \sigma_z/2$ which couples to the bath via $\hat{s}=\sigma_z$.
This is the \textit{independent boson model}~\cite{breuer_petruccione_2002} which we will use as an example here. Note however that what follows can be applied to any model which satisfies the commutation relation above.
\subsection{Independent Boson Model}
For this model the Trotter splitting in Eq.~\eqref{trot} is exact since the system and bath Hamiltonians commute.
Also, by moving to the interaction picture of the reduced system we may ignore the dynamics induced by $H_0$ and the free system propagator is therefore the identity
\begin{align}
\dbra{S_k}\text{e}^{\mathcal{L}_0 \Delta t}\dket{S_{k-1}} = \dbraket{S_k|S_{k-1}} \rangle=\delta_{s^+_ks^+_{k-1}}\delta_{s^-_ks^-_{k-1}}.
\end{align}
The summation in Eq.~\eqref{discreteevo} can now be carried out exactly (without the finite memory approximation) to obtain
\begin{align}
\label{indbossol}
\dbraket{S_N|\rho_R(t_N)} = \text{e}^{\Gamma(S_N,t_N)}\dbraket{S_N | \rho_0},
\end{align}
where
\begin{multline}
\Gamma(S_N,t_N)= -(s_N^+-s_N^-)^2 \Re[\eta(t_N)] \\ -i\left(\left({s_N^+}\right)^2 -\left({s_N^-}\right)^2\right)\Im[\eta(t_N)].
\end{multline}
Here we have defined the function
\begin{equation}
\label{lineshape}
\eta(t_N)=\sum_{k=0}^N \sum_{k'=0}^k\eta_{k-k'}=\int_0^{t_N}\int_0^{t'} C(t'-t'')dt''dt',
\end{equation}
which governs all dynamics induced by the interaction with the bath.
We see that state populations on the diagonal of $\rho_R(t)$ (where $s^+=s^-$) do not undergo evolution and all dynamics are in the decay and oscillations of the coherences.

In what follows we examine the effect of making the finite memory approximation on the function $\eta(t)$. Note that this also describes the effects of making this approximation on the discretized influence functional since the $\eta_{k-k'}$ coefficients can be written entirely in terms of this function,
\begin{equation}\label{eq:coeffseta}
\eta_{k-k'} = \begin{dcases*}
\eta(t_{k-k'+1})-2\eta(t_{k-k'})+\eta(t_{k-k'-1}) &\text{ $k\ne k'$}\\ 
\eta(t_1) &\text{ $k=k'$}
\end{dcases*}.
\end{equation}

To work out how the solution is changed when we impose the memory cutoff we must consider what happens to the function $\eta(t)$ when we remove  those coefficients for which $k-k'>\Delta k$ from the sum in Eq.~\eqref{lineshape}.
In Fig.~\ref{fig:memcut} we see there are two distinct regions of the integral domain left after performing the memory cutoff: when $t<\tau_c+\Delta t$, (region A) $\eta$ is unaffected by the cutoff and region B where $t>\tau_c+\Delta t$ and at least some correlations are cut off. 
The sum of the coefficients within region A is exactly $\eta(\tau_c+\Delta t)$, while using Eq.~\eqref{eq:coeffseta} the sum of all the coefficients in a single strip of region B is $\eta(\tau_c+\Delta t)-\eta(\tau_c)$.
There are $N-\Delta k-1$ of these strips in total and so by summing up all coefficients remaining after the finite memory approximation we find the function $\eta(t)$ is approximated as
\begin{align}
\eta(t)\approx \eta(\tau_c)+(t-\tau_c)\frac{\eta(\tau_c+\Delta t)-\eta(\tau_c)}{\Delta t}.
\end{align}
This result can also be obtained in the $\Delta t \to 0$ limit by setting $C(t)=0$ for $t>\tau_c$ in the integral of Eq.~\eqref{lineshape}. By making a change of variable $s=t'-t''$, we find
\begin{align}
\eta(t)=&\int_0^t\int_0^{t'} C(s)dsdt'\\
 \approx&  \int_0^{\tau_c} \int_0^{t'}C(s)ds dt' +\int_{\tau_c}^t \int_{0}^{\tau_c}C(s)dsdt'\\
=&\eta(\tau_c)+(t-\tau_c)\dot{\eta}(\tau_c).
\end{align}
where the dot indicates a time derivative.
Thus the dynamics of the pure dephasing model with the finite memory approximation imposed are as follows:
At times $t\leq\tau_c+\Delta t$ the exact dynamics are followed. While for $t>\tau_c+ \Delta t$ there is an exponential decay with rate $\gamma_c=\Re[\dot{\eta}(\tau_c)]$ accompanied by a Lamb-shift of the system frequencies, $\Lambda_c=\Im[\dot{\eta}(\tau_c)]$.

It is easy to see now how a problem can arise if the memory cutoff is at a point where the gradient of $\eta(t)$ is negative. The resultant decay rate is negative and hence there is exponential growth of all coherences for $t>\tau_c$.
This inevitably leads to the system density matrix becoming non-positive.

Additionally, unless $\Re[\dot{\eta}(\tau_c)]=0$ identically, finite steady state coherences will be impossible to produce.
Increasing $\tau_c$ increases the time it takes for the coherences to decay or grow away from the true steady state, but for finite $\tau_c$ the convergence of the finite memory approximation, and hence the ADT scheme, cannot be guaranteed at arbitrary times.

\begin{figure}
\includegraphics[scale=0.4]{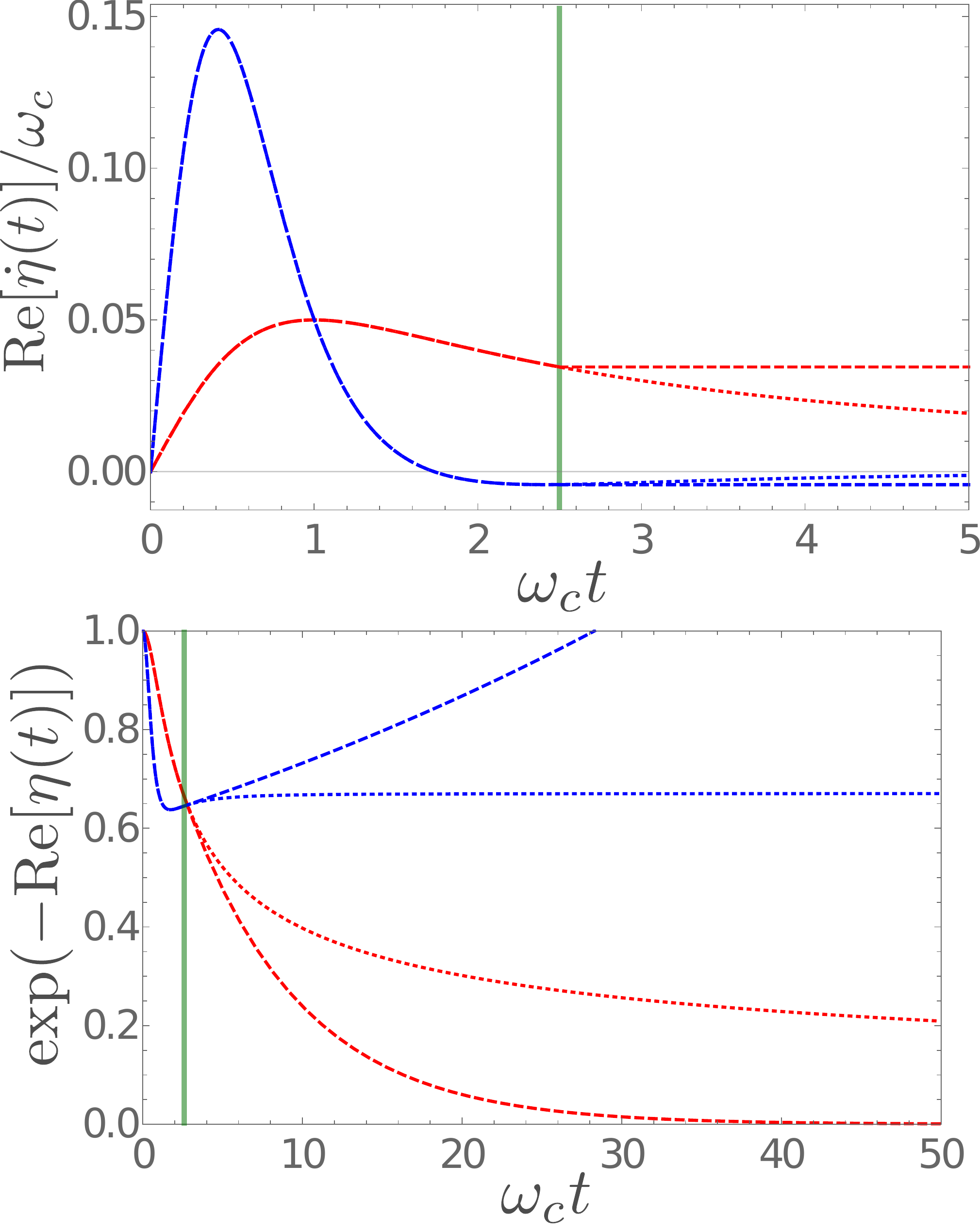}
\caption{ The decay rate $\dot{\eta}(t)$ and dynamics $\exp(-4\Re[\eta(t)])$ for $\nu=1$, $\nu=3$, in red and blue respectively, for exact results (dotted) and with the memory cutoff (dashed).
The vertical lines indicate the memory cutoff time $\omega_c\tau_c=2.5$ and other parameters are $T=0$ and $a=0.2$.
With the memory cutoff the decay rate is exact until $\tau_c$, after which it forever assumes its value at $\tau_c$. 
}
\label{fig:quapr}
\end{figure}

To investigate how these problems manifest in the dynamics of a real example we need to specify the environmental coupling, quantified by the spectral density $J(\omega)$, which determines the form of $\eta(t)$.
We consider the spectral density
\begin{equation}\label{specd}
J(\omega)=\frac{\alpha}{2} \frac{\omega^\nu}{\omega_c^{\nu-1}}\text{e}^{-\omega/\omega_c},
\end{equation}
which is characterised by the dimensionless coupling constant, $\alpha$, ohmicity, $\nu$ and an exponential cutoff at a scale given by $\omega_c$.
It has been found that the general condition for $\dot{\eta}(t)>0$ is that the function $J(\omega)\coth(\omega/2T)/\omega^2$ be convex, and that for spectral densities of the form Eq.~\ref{specd} this condition is fulfilled when $\nu<\nu_{\text{crit}}$ for some critical ohmicity $\nu_{\text{crit}}$~\cite{haikka_johnson_maniscalco_2013}.
The value of $\nu_{\text{crit}}$ varies monotonically from $\nu_{\text{crit}}=2$  and $\nu_{\text{crit}}=3$ as temperature is increased from 0 to $\infty$. 
Negative decay rates arise when $\nu>\nu_{\text{crit}}$.
Therefore at $T=0$ it is in the superohmic regime, where $\nu>2$, that we expect to find the finite memory cutoff results in non-positive dynamics.
To demonstrate this we plot in Fig.~\ref{fig:quapr} the time dependent decay rate resulting from both ohmic and superohmic spectral densities at zero temperature.
We also plot the resultant dynamics of the coherence, for both the exact solution of the independent boson model, Eq.~\eqref{indbossol}, and with the memory cutoff imposed.
In this case the imaginary part of the exponential in Eq.~\eqref{indbossol} disappears, since the eigenvalues of $\hat{s}$ have equal magnitude, so the dynamics are that of pure decay.
For the ohmic case the decay rate is always positive and using the memory cutoff produces exponential decay to the correct steady-state of zero coherence. There is still a significant discrepancy at intermediate times, but this can be reduced by increasing $\tau
_c$.
For a superohmic spectral density there is a large window of time when the decay is negative. Making the cutoff in this window results in unphysical dynamics with exponential growth.
Notably, in this superohmic case the decay rate approaches zero from below asymptotically. This means that increasing $\tau_c$  does not remove this spurious asymptotic behaviour it just shifts its onset to longer and longer times.

To conclude this analysis we point out a situation in which even the ohmic and sub-ohmic regimes $0<\nu\leq 2$ can display negative decay rates and hence non-positive states.
In constructing physically realistic models one may want to account for how spatially separated states interact with the same bath of oscillators.
For the case of the two-level independent boson model considered above in three dimensions the spatial separation of the two system eigenstates is accounted for via a multiplicative correction to the spectral density $\propto 1-\sinc(k\omega )$~\cite{mccutcheon11}.
Here $k=d/c$ where $d$ is the distance between the two sites and $c$ is the speed at which the bath excitations propagate.
Thus, for small separation distance, $k\omega_c<1$, spectral densities of the form Eq.~\eqref{specd} go as $J(\omega)\propto \omega^{\nu+2}e^{-\omega /\omega_c}$ and the effective critical ohmicity for the transition to non-Markovian dynamics with negative decay rates now lies in the range $0<\nu_{crit}\leq 1$ and the potential for the memory cutoff to produce unphysical dynamics exists for all values of $\nu$.

\subsection{Fixing the non-positive evolution}

\begin{figure}
\includegraphics[scale=0.7]{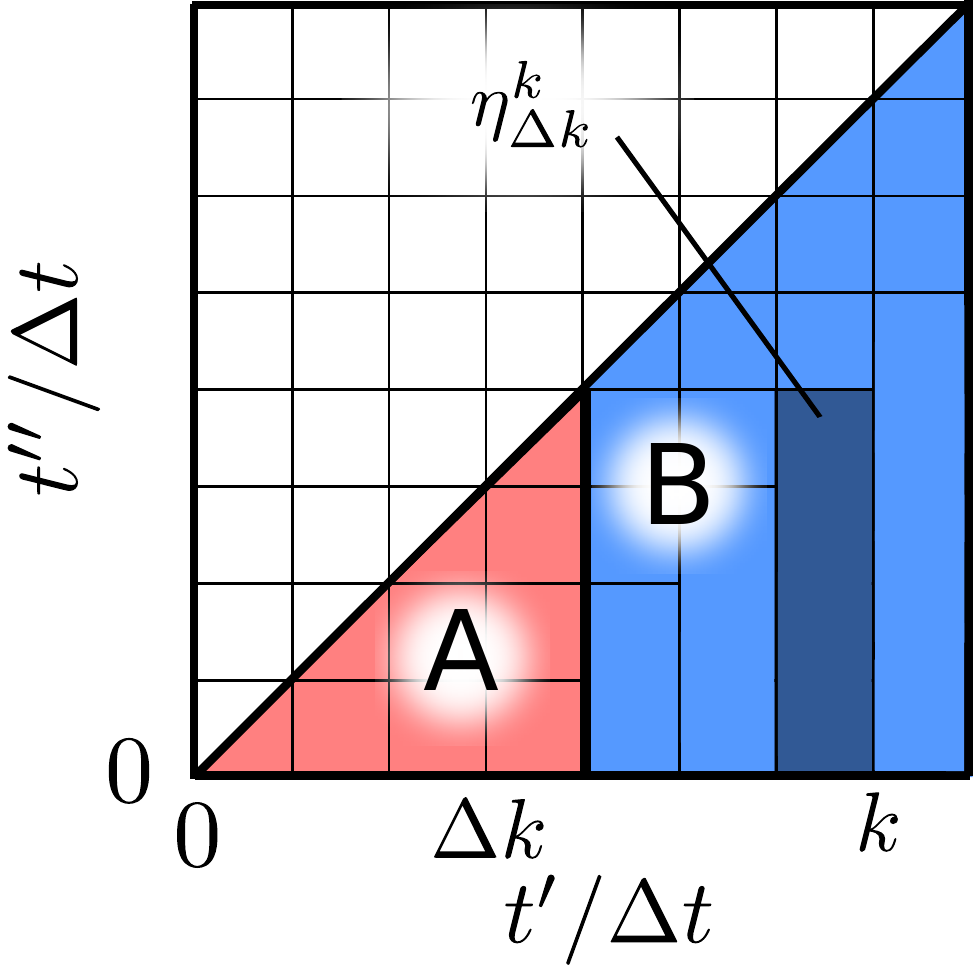}
\caption{Visualisation of the modified memory cutoff approximation.
Instead of discarding the regions of the integral domain which correspond to $\eta_{k-k'}$  coefficients giving correlations over larger times than the memory cutoff, the integral domains for the coefficients at the ``edge" of the memory cutoff are extended from squares to rectangles that reach back to zero.
In this case the full integral domain is maintained.}
\label{fig:newcoef}
\end{figure}

We now seek a way of taking the memory cutoff such that the exact solution Eq.~\ref{indbossol} is always reproduced using the ADT method.
A general way to do this would be to define a new set of the coefficients $\tilde{\eta}_{k-k'}$, $k-k'\leq \Delta k$ in such a way that their sum is constrained as follows,
\begin{align}
\sum_{k=0}^N\sum_{ k'=k-\Delta k}^{k} \tilde{\eta}_{k- k'}=\sum_{k=0}^N\sum_{k'=0}^k \eta_{k- k'}=\eta(t),
\end{align}
thus reproducing the exact dynamics governed by $\eta(t)$, independent of both $\tau_c$ and $\Delta t$.
It seems reasonable that we should attempt to redefine as few of the coefficients as possible, since they already exactly account for non-Markovian correlations within a timespan $\tau_c$ (up to the Trotter error).
The key idea behind the finite memory cutoff is that the most non-local temporal correlations in the system are the most insignificant, hence $|\eta_{k-k'}|\approx 0$ for large $k-k'$.
Thus we redefine only the coefficients with the largest $k-k'=\Delta k$ as follows
\begin{equation}\label{eq:newcoeffs}
\tilde{\eta}_{k-k'}= \begin{dcases*}
\eta_{k- k'}+\sum_{j=0}^{k-\Delta k-1} \eta_{k-j}\equiv \eta_{\Delta k}^k  &\text{ $k- k'=\Delta k$}\\ 
\eta_{k- k'} &\text{ $k- k'<\Delta k$}
\end{dcases*}.
\end{equation}
This redefinition will not introduce any problems as long as $|\eta_{k-k'}|\approx|\tilde{\eta}_{k-k'}|$ which is true since $\tau_c$ is already large enough to na\"ively justify the finite memory approximation.
This redefinition is visualized in Fig.~\ref{fig:newcoef} in a similar way to the standard memory cutoff in Fig.~\ref{fig:memcut}.
From this schematic one can identify that this redefinition corresponds to extending the lower limit of the integral down to zero. This means that it is straightforward to calculate the redefined coefficient:
\begin{align} 
\eta^k_{\Delta k}&=\int_{t_{k-1}}^{t_k}\int_{0}^{t_k - \tau_c}C(t'-t'') dt'' dt' \nonumber\\
&=\eta(t_{k})-\eta(t_{k-1})-\eta(\tau_c)+\eta(\tau_c-\Delta t).
\end{align}
In terms of implementing the ADT scheme the main consequence is that now a new propagator must be constructed at every step of the iterative propagation, though the actual structure of each propagator is essentially identical to the original. Thus, although slightly more time consuming, carrying out the propagation is no more complicated than before.
The time dependence in our method is fundamentally different from that generated by time-dependent Hamiltonians, in that it is due to the non-local influence of the bath interaction itself.
In the standard finite memory approximation the finite ranged non-Markovian correlations allow the problem to be reformulated as one with Markovian evolution but in a higher dimensional augmented Hilbert space as in Eq. (\ref{eq:quapiprop}).
We have gone one step further by allowing $\mathcal{L}_A$ to be time dependent, thus allowing transient non-positive evolution of the reduced system state beyond the cutoff time $\tau_c$ while still maintaining overall positive evolution from the initial state. This is not possible in the standard ADT scheme where the augmented Liouvillian is time independent and so  must have no positive eigenvalues to ensure positive evolution. We point out here that the instability of steady states and non-positivity in the ADT scheme for superohmic environments has been recognized before, and a similar redefinition of the $\eta_{k-k'}$ coefficients has been proposed~\cite{quapi_so1, quapi_so2} but only in a way that gives qualitatively correct asymptotic behaviour of the pure dephasing model above, but does not reproduce the exact solution identically.

In the following section we show that this new method is useful for more complicated models which do not have an exact solution by applying it to the more general spin-boson model.

\section {Application to the Spin-Boson Model}
\label{sec:spinbos}

The spin-boson model~\cite{leggett87} provides a paradigmatic example of an open quantum system and can be used to model a wide variety of physical systems. For example, it has been applied to the problem of electron transfer in biological aggregates~\cite{xu94}, exciton dynamics in quantum dots~\cite{QDdamping}, transport in mesoscopic systems~\cite{brandes05} and chemical reactions~\cite{garg85}. The system Hamiltonian is $H_0=\epsilon \sigma_z + V\sigma_x$ and the coupling to the environment is through the operator $\hat{s}=\sigma_z$. Here $\epsilon$ again gives the energy splitting of the two levels, while the $\sigma_x$ term generates coherent transitions between the two states. The addition of this $\sigma_x$ term breaks the integrability of the independent boson model and allows for a rich variety of physics to be explored.

In what follows we will be particularly interested in the case where the environment is superohmic, since this is where we found the most pathological behaviour in the independent boson model. The superohmic regime is most studied in the context of quantum dots strongly coupled to a phononic environment~\cite{QDlambshift, QDdamping, kok_lovett_2010}. For many parameters a polaron master equation provides a successful route to capturing non-perturbative effects~\cite{nazir_mccutcheon_2016}, but for highly non-Markovian environments this approach fails. Here we show how the ADT scheme is able to capture backflow of energy from the environment to the system in this regime.

\begin{figure}
\includegraphics[width=\columnwidth]{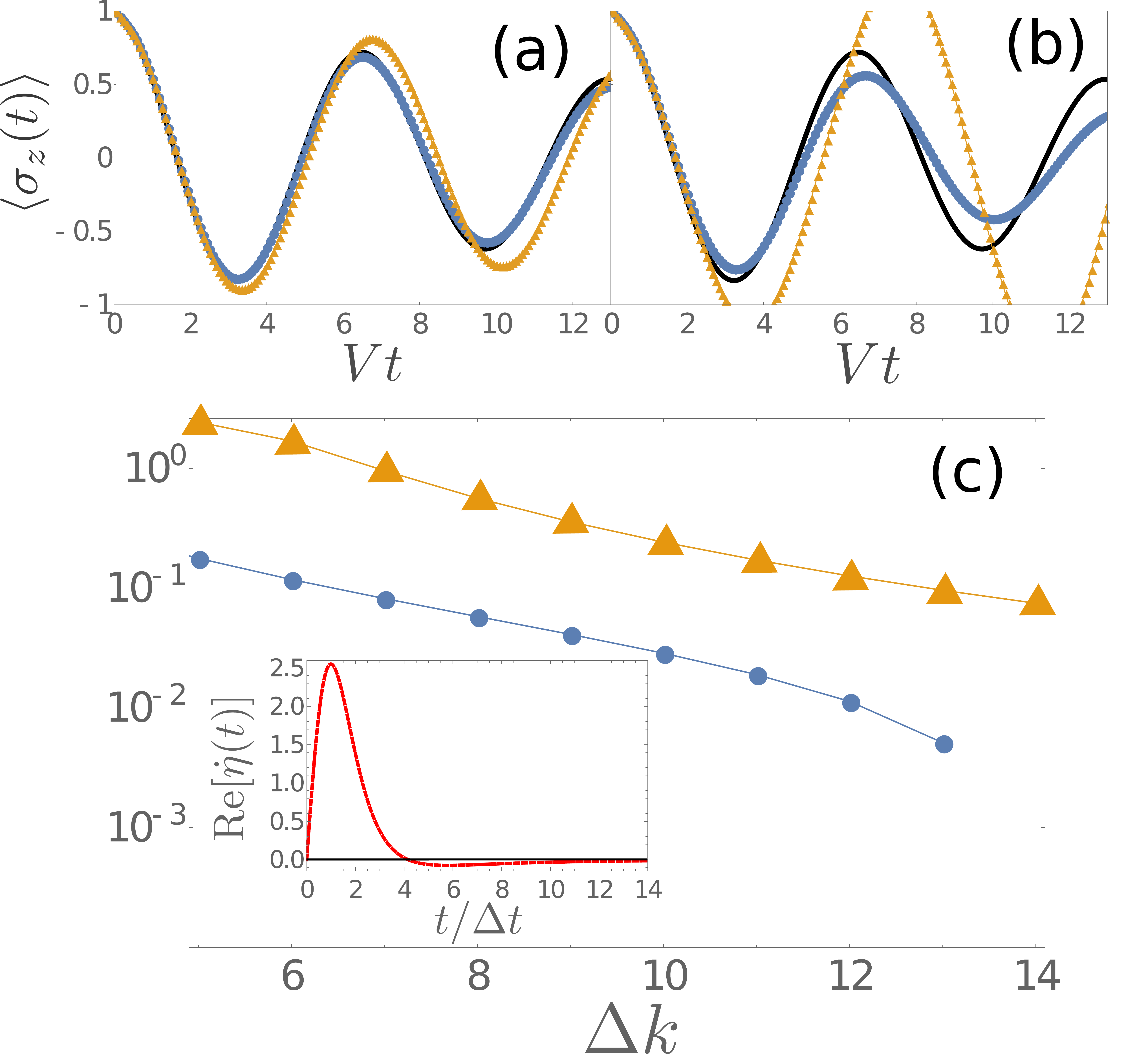}
\caption{Top panels: Dynamics of the spin-boson model for (a) $\Delta k=11$  and (b) $\Delta k = 6$ compared to the converged results (black) for the improved memory approximation (blue) and the standard ADT (orange). Panel (c): The trace distance between the converged result against the value of $\Delta k$ used for the standard finite memory approximation (orange triangles) and our improved memory approximation  (blue dots). The inset shows the time dependent decay function. Parameters in all panels are $\alpha=0.7$, $\omega_c/V=5$, $T=0$.}
\label{fig:convplot}
\end{figure}

To show the improvement gained by the new integration scheme which we detailed in the previous section, we show how the convergence of the results with $\Delta k$ changes as compared to the standard approach.
In Fig.~\ref{fig:convplot} we study the spin-boson model with $\epsilon=0$ at zero temperature with a large reservoir cutoff frequency $\omega_c/V=5$. To find an approximation for the error in our results we compare everything to the most converged case: that using the new memory cutoff at $\Delta k =14$. We then calculate the time average of the trace distance~\cite{breuercoll} between these converged results and each other case. The trace distance is given by
\begin{align}
D(\rho_1(t),\rho_2(t))=\frac{1}{2} \Tr |\rho_1(t)-\rho_2(t)|,
\end{align}
where where $|A|=\sqrt{AA^\dagger}$, $\rho_1(t)$ is the converged density matrix and $\rho_2(t)$ is the density matrix we want to compare to $\rho_1(t)$.
We show this deviation as a function of $\Delta k$ for both approaches in  the panel (c) of Fig.~\ref{fig:convplot}. We see that our new approach converges much more quickly than the standard finite memory approximation: by $\Delta k=11$ the errors in the new approach are negligible while the standard finite memory approximation still has significant errors at $\Delta k=14$. This can also be seen in the dynamics in Fig~\ref{fig:convplot}(a) and (b) where we see that our new approach converges much more quickly when increasing $\Delta k$.

It is difficult to find convergence with this set of parameters because the gradient of $\eta(t)$ approaches zero from below (as can be seen in the inset to Fig.~\ref{fig:convplot}(c)) and so the problems we found for the independent boson model still occur, although here they are less severe and the standard approach only gives unphysical results at very small $\Delta k$. 
This problem becomes even worse if we move further towards the so-called scaling limit of the model~\cite{leggett87}, where $\omega_c$ is the largest energy scale in the problem, by increasing the value of $\omega_c$. This means a smaller required $\tau_c$, but also results in a larger magnitude of $\dot{\eta}(\tau_c)$ so that the timescale for the onset of divergent dynamics becomes much shorter and achieving convergence over appreciable lengths of time becomes very difficult.

\begin{figure}
\includegraphics[scale=0.25]{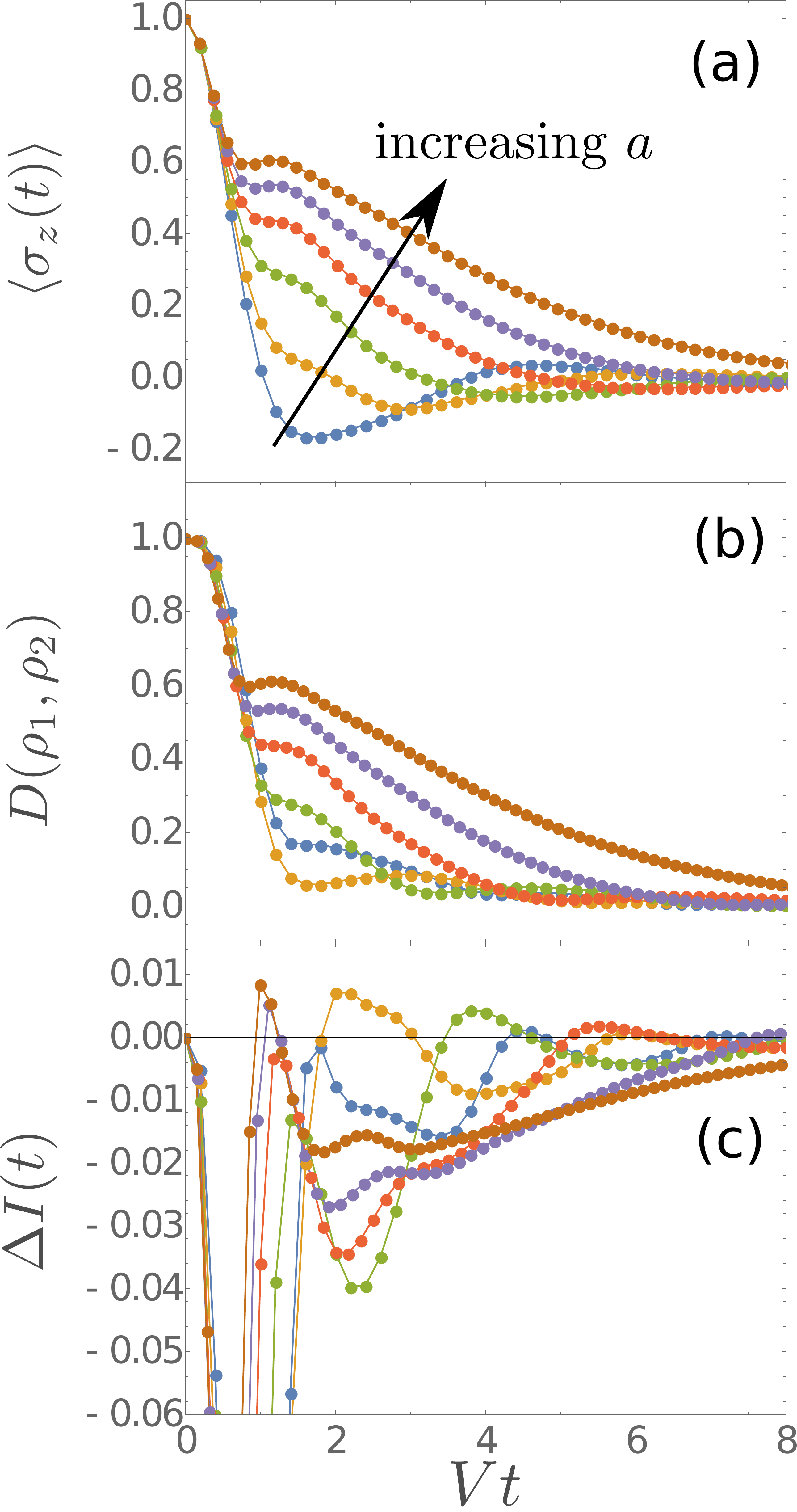}
\caption{(a) Population difference dynamics for various values of coupling strength to the bath $\alpha=0.5,0.7,1,1.3,1.6,1.9$. (b) Trace distance dynamics for 2 orthogonal initial conditions for the same parameters as (a). (c) The derivative of the trace distance showing regions where the dynamics is non-Markovian.
Other parameters are $\omega_c/V=T/V=1$.}
\label{fig:bumpplot}
\end{figure}

Now we have shown the improvement gained by our improved memory cutoff we examine in detail the dynamics of the spin-boson model in a parameter regime which is difficult to access using the standard memory cutoff approximation: when there is non-Markovian behaviour far from the scaling limit with a lower cutoff frequency $\omega_c/V=1$ and at finite temperature. In this limit the required value of $\Delta k$ without our improvement would be much too large to store in memory for a standard computer.

In Fig.~\ref{fig:bumpplot}(a) we show the population difference dynamics with initially excited spin, $\rho_0=\ket{1}\bra{1}$, for system-bath couplings between $0.5<\alpha<1.9$.
At low coupling strengths we see simple damped oscillations, as would be captured by a weak coupling master equation.
Increasing the coupling changes these oscillations from underdamped to overdamped, as would be found using for example a polaron master equation. However, the overdamped regime is accompanied by a highly non-Markovian feature: a revival of the population difference which becomes more pronounced at stronger couplings.

In order to establish a more concrete quantification of the non-Markovian nature of these dynamics  we examine how distinguishable two distinct initial states are as they evolve in time. To do this we again use the trace distance $D(\rho_1(t),\rho_2(t))$ as a natural measure.
For a Markovian system as time evolves all states move closer to the steady state and so asymptotically we expect $\lim_{t \to \infty} \rho_1(t)=\lim_{t \to \infty} \rho_2(t)=\rho_{\text{steady}}$, provided the steady state $\rho_{\text{steady}}$ is unique. This means that $D(\rho_1(t),\rho_2(t))$ monotonically decays to 0 for any two initial states. 
If, however, the dynamics are non-Markovian then, for certain pairs of initial states, there can be increases in the trace distance as a function of time as the environment allows information to flow back into the system.
It is therefore evident that quantity of interest is actually
\begin{equation}
\Delta I(t)=\frac{d}{dt}D(\rho_1(t),\rho_2(t)),
\end{equation}
which is positive when these non-Markovian features occur.

In Fig.~\ref{fig:bumpplot}(b) and (c) we show the trace distance and $\Delta I(t)$, for the same set of parameters as in (a), using the orthogonal initial states $\rho_1(t)=\ket{1}\bra{1}$ and $\rho_2(t)=\ket{-1}\bra{-1}$.
At low values of coupling there are weak oscillations in $\Delta I(t)$ between negative and positive which are damped and pushed out to longer times as the coupling is increased.
Increasing the coupling still further leads to an additional positive region of $\Delta I(t)$ at short times, which remains there, closely following the early time revival we see in the population differences in Fig.~\ref{fig:bumpplot}. This observation confirms that this feature is a signature of information backflow. Similar features have been observed in biased spin-boson models (i.e. $\epsilon\neq 0$) \cite{chin10,wilner_wang_thoss_rabani_2015,jang2009theory}. Here, we show quantitatively that non-Markovian features are present in the dynamics of the unbiased spin-boson model.

\begin{figure}
\includegraphics[width=0.45\textwidth]{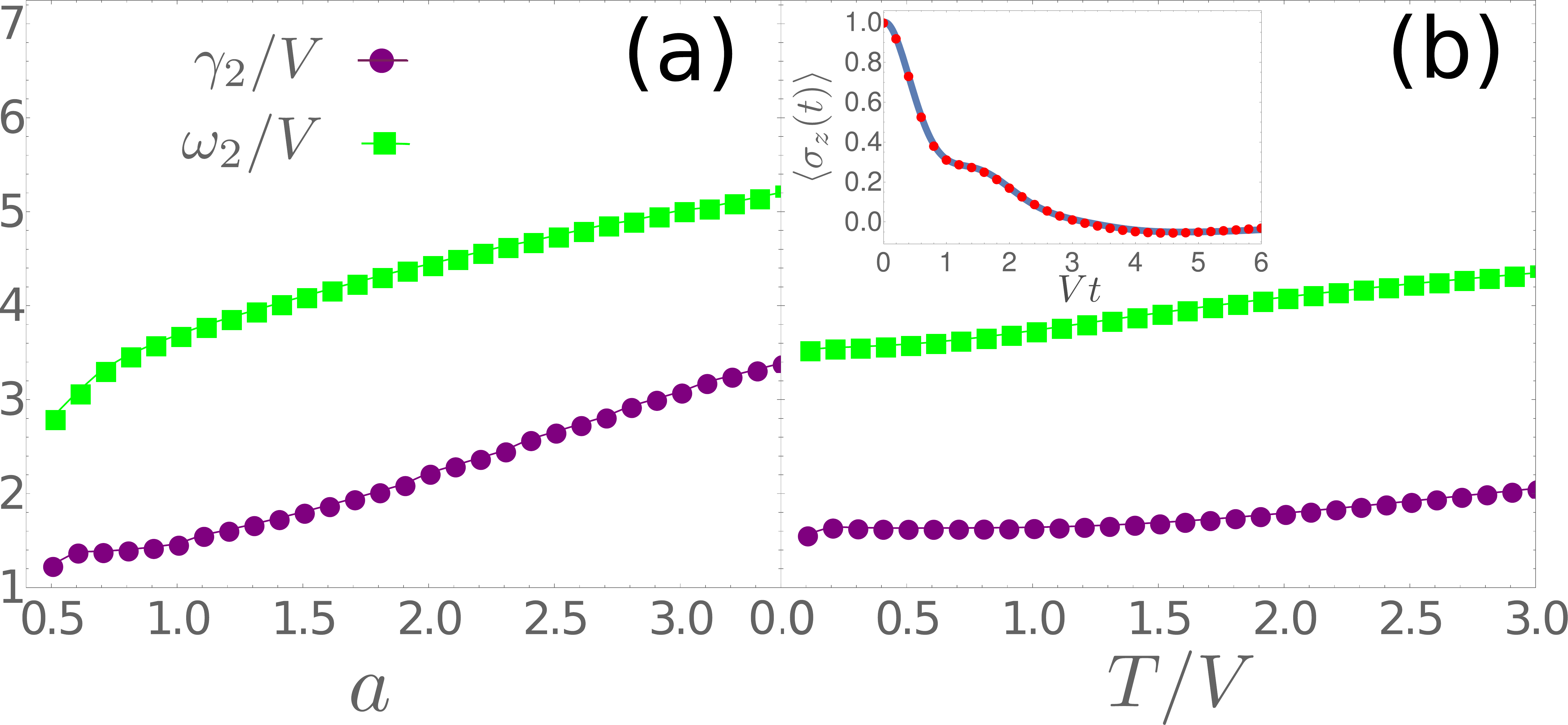}
\caption{The decay $\gamma_2/V$ (purple dots) and frequency $\omega_2/V$ (green squares) fitting parameters corresponding to the high frequency sinusoidal oscillation present on top of the decaying cosine. In (a) these are plotted against dimensionless coupling $\alpha$ with $\omega_c/V=T/V=1$. In (b) these are plotted against temperature  with $\omega_c/V=\alpha=1$. The inset shows an example of the fit to Eq.~\ref{fit} (blue solid line) compared to the numerical results (red dots) at  $\alpha=1$.}
\label{fig:pcouplot} 
\end{figure}

To analyse the dependence of the non-Markovian revival in the population dynamics on the parameters describing the bath we fit the dynamics to the sum of two decaying oscillations:
\begin{align}\label{fit}
\alpha_z(t)=Ae^{-\gamma_1 t}\cos(\omega_1 t)+Be^{-\gamma_2 t}\sin(\omega_2 t+\phi),
\end{align}
with the constraint $\alpha_z(0)=A+B\sin(\phi)=1$. 
The frequency $\omega_1$ and decay rate $\gamma_1$ parameters capture the weak-coupling and long time dynamics, while $\omega_2$ and $\gamma_2$ describe the higher frequency short time dynamics of the revival. An example of one of these fits, for $\alpha=0.6$, is shown in the inset to Fig.~\ref{fig:pcouplot}.

We show how $\gamma_2/V$ and $\omega_2/V$ vary as the coupling to the bath, $\alpha$, is increased in Fig.~\ref{fig:pcouplot}(a).
Both $\gamma_2/V$ and $\omega_2/V$ increase with coupling at roughly the same rate and, although the behaviour is certainly not linear, this dependence implies that increasing coupling simply causes the timescale over which the revival occurs to shorten.
This is consistent with the revival being due to a backflow of information from the bath: we would expect larger couplings to simultaneously cause quicker information backflow and to wash it out more quickly. In Fig.~\ref{fig:pcouplot}(b) we show what happens as the temperature of the bath is changed.  Increasing temperature both increases the effective coupling to the bath but also scrambles temporal correlations, reducing the bath correlation time. Therefore we see both the damping and frequency of the revival increase with increasing temperature.

\begin{figure}
\includegraphics[width=0.49\textwidth]{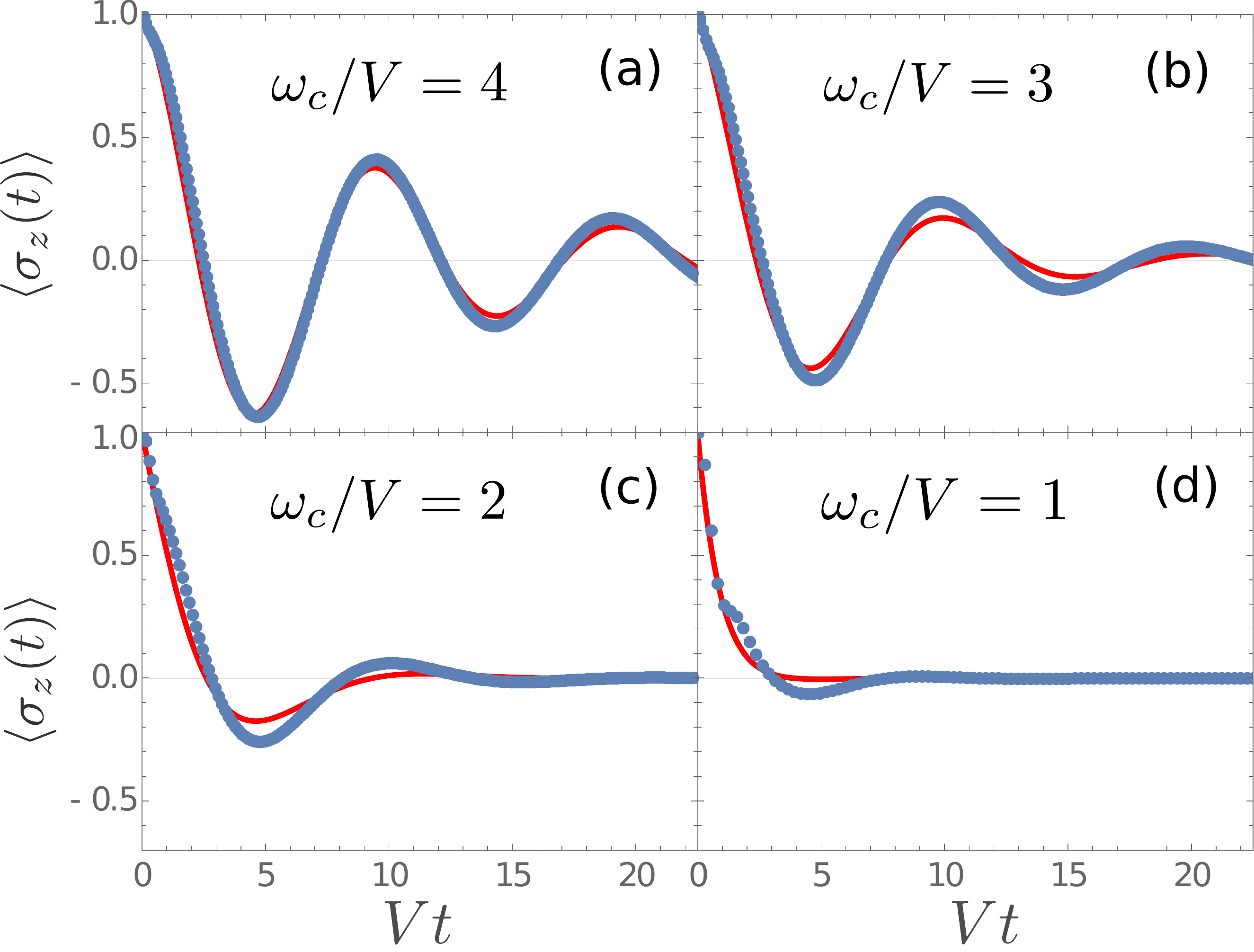}
\caption{Population difference plotted using the ADT with our improved memory cutoff (blue dots) and the polaron master equation (red solid) for (a) $\omega_c/V=4$, (b) $\omega_c/V=3$, (c) $\omega_c/V=2$, (d) $\omega_c/V=1$. 
Other parameters used were $\alpha=1$ and $T/V=1$.  }
\label{fig:polaroncomp}
\end{figure}

Before concluding, we compare the results above to those obtained using a Markovian polaron master equation~\cite{nazir_mccutcheon_2016}.
If our interpretation of the revival is correct, then it is obvious that any approach which does not fully account for the dynamics of the bath will be unable to reproduce this feature. For example, the population difference dynamics predicted using the polaron master equation are of the same form as Eq.~\eqref{fit} but with $\gamma_1=\gamma_2$ and $\omega_1=\omega_2$~\cite{mccutcheon11, nazir_mccutcheon_2016}.
This is shown in Fig.~\ref{fig:polaroncomp}. 
For the largest values of $\omega_c$ in (a) and (b)  the system is in the scaling limit~\cite{leggett87}, where the polaron technique works well, and indeed we find good agreement of the two methods, with only small deviations occurring at very short timescales.
As $\omega_c$ decreases, we enter a more non-Markovian regime and we see that the polaron master equation starts to fail. 
It is difficult to find converged results which show this effect more clearly than shown here, since in the scaling limit divergent dynamics occur on very short timescales, as discussed earlier.
The long time dynamics predicted by the polaron equation are qualitatively correct, taking the form of an exponentially decaying oscillation, but as anticipated it completely fails to capture the non-Markovian revival which we found above using the improved ADT.
We attribute this to the breakdown of the Markov approximation used in deriving the polaron master equation.
From our earlier discussion, we know that the revival in dynamics is most prominent for smaller values of $\omega_c$, i.e.\ for baths with longer correlation times.
This, together with the poor agreement of ADT with the polaron master equation at short times, confirms our conclusion that this feature is a highly non-Markovian effect which could not be produced using any form of time-local master equation. 

\section{Summary}
\label{sec:summary}

We have shown how the standard finite memory approximation in the ADT numerical scheme can cause unphysical behaviour resulting in periods of non-positive evolution. We have provided an improvement to this method which is able to reproduce the exact solution to the pure dephasng model.
To demonstrate the applicability of this new method we have shown converged results for dynamics of the symmetric spin-boson model in a superohmic environment. We found that our new method is able to reach convergence using significantly smaller computational resources than the standard finite memory approximation.
At strong system-bath coupling strengths in this regime we found highly non-Markovian revivals in the dynamics which are accompanied by non-monotonicity of the trace distance between different initial states. These features are not present in simpler analytical approaches such as polaron master equations.

\section*{Acknowledgements}

We acknowledge useful discussions with E.~M.~Gauger. We thank D.~P.~S.~McCutcheon for providing code used for the polaron simulations. AS acknowledges a studentship from EPSRC (EP/L505079/1). BWL acknowledges support from EPSRC (EP/K025562/1).
PGK acknowledges support from EPSRC (EP/M010910/1). The research data supporting this publication can be found at \url{http://dx.doi.org/10.17630/21764101-6493-46e7-ab85-65cccc1fb9e5}.

\appendix

\section{Explicit Construction of Propagation Methods}
\label{app:prop}
We start by rewriting the summand of Eq.~\eqref{discreteevo} in the following form
\begin{multline}
F\left(\{S_k\}\right) I\left(\{S_k\}\right)=\left(\prod_{j=1}^N \prod_{j'=1}^j \tilde{I}(S_j,S_{j'})\right) \\\times\dbra{S_1}e^{\mathcal{L}_0 \Delta t}\dket{S_0}
\end{multline}
where
\begin{equation}
\tilde{I}(S_j,S_{j'}) = \begin{dcases*}
e^{- \phi(S_j,S_{j'})} &\text{ $j-j'\ne 1$}\\ 
\dbra{S_j}e^{\mathcal{L}_0 \Delta t}\dket{S_{j'}}e^{- \phi(S_j,S_{j'})} &\text{ $j-j'=1$}
\end{dcases*},
\end{equation}
with
\begin{equation}
 \phi(S_j,S_{j'})=(s^+_j-s^-_j)(s^+_{j'}\eta_{j-j'}-s^-_{j'}\eta_{j-j'}^*).
\end{equation}
There is no $S_0$ dependence in any of the $\tilde{I}(S_k,S_{k'})$ so the $S_0$ summation in Eq.~\eqref{discreteevo} can be carried out, propagating the reduced system for a time $\Delta t$ under the free Liouvillian $\mathcal{L}_0$ into the state $\rho_R(\Delta t)$. 
The finite memory approximation, setting $\eta_{k-k'}=0$ for $k-k'>\Delta k$, then translates to $\tilde{I}(S_k,S_{k'})=1$ for $k-k'>\Delta k$. This allows the product over $\tilde{I}(S_k,S_{k'})$'s to be rewritten as
\begin{multline}
 \prod_{j=1}^N \prod_{j'=1}^j \tilde{I}(S_j,S_{j'}) \approx \prod_{j=\Delta k +1}^N \Lambda(S_j, S_{j-1} \ldots S_{j-\Delta k})\\
 \times A(S_{\Delta k},S_{\Delta k-1} \ldots S_1),
\end{multline}
where we have defined
\begin{equation}\label{lambda}
\Lambda(S_j, S_{j-1} \ldots S_{j-\Delta k})=\prod_{m=0}^{\Delta k}\tilde{I}(S_j,S_{j-m}),
\end{equation}
to be the elements of the propagator tensor, and
\begin{equation}\label{initaug}
A(S_{\Delta k}, S_{\Delta k-1} \ldots S_{1})=\dbraket{S_1|\rho_R(\Delta t)} \prod_{k=1}^{\Delta k}\prod_{k'=1}^{ k}\tilde{I}(S_k,S_{k'}),
\end{equation}
to be the elements of the initial augmented density tensor.
The summations over the $S_k$ can now be carried out one at a time by iteratively multiplying and contracting tensors:
\begin{multline}
A(S_{k}, S_{k-1} \ldots S_{k-\Delta k+1})=\sum_{S_{k-\Delta k}} \Lambda(S_k, S_{k-1} \ldots S_{k-\Delta k})\\
\times A(S_{k-1}, S_{k-2} \ldots S_{k-\Delta k}),
\end{multline}
with the initial condition Eq.~\eqref{initaug}.
The reduced system density matrix at a time $t_k$ is then retrieved by summing over all but the $S_k$ index
\begin{equation}
 \dbraket{S_k|\rho_R(t_k)}=\sum_{S_{k-1}\ldots S_{k-\Delta k +1}}A(S_{k}, S_{k-1} \ldots S_{k-\Delta k+1}).
\end{equation}
The propagator tensor Eq.~\eqref{lambda} has $\Delta k+1$  indices and so has $D^{2(\Delta k+1)}$ elements, making the tensor contraction representation of the propagation less demanding than representing the augmented state as a vector and propagating with a $D^{4 \Delta k}$ sized matrix.
Within the Liouvillian matrix formalism for propagation, Eq.~\eqref{eq:quapiprop}, and using the basis defined in Eq.~\eqref{eq:lbasis}, the elements of the augmented state at a time $t_k$ are:
\begin{equation}
\dbraket{S^A_k|\rSA(t_k)}=A(S_{k}, S_{k-1} \ldots S_{k-\Delta k+1}),
\end{equation}
and the matrix elements of the propagator across the timespan $\tau_c$ have the simple analytic form:
\begin{equation}
 \dbra{S_k^A}e^{\mathcal{L}^A \tau_c}\dket{S_{k-\Delta k}^A}=\prod_{j=k-\Delta k+1}^k \Lambda(S_j, S_{j-1} \ldots S_{j-\Delta k}).
\end{equation}
Thus another way of solving the problem would be to construct and diagonalize this propagator to find the eigenvectors of the augmented Liouvillian $\mathcal{L}^A$, though for typical values $\Delta k \sim 10$ it is much more efficient to use the iterative propagation schemes above.

\end{document}